\documentclass[letter,twocolumn]{jpsj3}
\usepackage{txfonts}

\usepackage{braket}
\usepackage{bm}
\usepackage{graphicx}
\usepackage{dcolumn}
\usepackage{bm}
\usepackage{amsmath}
\usepackage{times}
\usepackage[usenames]{color}
\usepackage{ulem}

\newcommand{\kets}[1]{|#1\rangle}

\newcommand{\means}[1]{\langle#1\rangle}

\def\journal #1#2#3#4{#1 {\bf #2}, #3 (#4)}

\def\PRB{Phys.\ Rev.\ B}

\def\JPSJ{J.\ Phys.\ Soc.\ Jpn.}

\def\RMP{Rev.\ Mod.\ Phys.}

\title{Magneto-Electric Effect in a Spin-State Transition System}

\author{Makoto Naka$^{1}$\thanks{naka@aoni.waseda.jp}, Eriko Mizoguchi$^{2}$, Joji Nasu$^{3}$,  and Sumio  Ishihara$^{2}$}
\inst{
$^1$Waseda Institute for Advanced Study, Waseda University, Tokyo 169-8050, Japan \\
$^2$Department of Physics, Tohoku University, Sendai 980-8578, Japan \\
$^3$Department of Physics, Tokyo Institute of Technology, Meguro, Tokyo 152- 8551, Japan
}

\abst{
Magnetic, dielectric, and magnetoelectric properties in a spin-state transition system are examined, motivated by the recent discovery of a multiferroic behavior in a cobalt oxide. 
We construct an effective model Hamiltonian based on the two-orbital Hubbard model, in which the spin-state degrees of freedom in magnetic ions couple with ferroelectric-type lattice distortions. 
A phase transition occurs from the high-temperature low-spin phase to the low-temperature high-spin ferroelectric phase with accompanying an increase of the spin entropy. 
The calculated results are consistent with the experimental pressure-temperature phase diagram. 
We predict the magnetic-field induced electric polarization in the low-spin paraelectric phase near the ferroelectric phase boundary. 
}

\begin{document}
\maketitle

Multiferroics are the coexistence phenomena of the ferromagnetic, ferroelectric, and ferroelastic orders.~\cite{schmid} 
The cross-correlation effects between the magnetism and dielectrics are termed the magnetoelectric (ME) effect.~\cite{curie, dzyaloshinskii, astrov, folen, tachiki, fiebig} 
These issues have attracted recently much attention not only from the fundamental condensed matter physics, but also the wide potentiality for the technological applications. 
One of the predominant target materials of the multiferroics and the ME effect is the Mott insulating systems, in which the electronic charge degree of freedom is quenched and the spin degree of freedom dominates the low energy physics. 
The well studied examples are the systems showing non-collinear and non-coplanar spin orders owing to the competing exchange interactions.~\cite{shiratori, kimura} 
The spontaneous electric polarizations are induced by the so-called inverse Dzyaloshinsky-Moriya interaction mechanism, and are controlled by an external magnetic field.~\cite{katsura, dagotto, mostovoy} 
Now, the candidates of the multiferroics and ME effect are surveyed extensively in a wide variety of materials.~\cite{n_ikeda, khomskii, kimura2, arima, naka} 

In some classes of the magnetic ions, multiple spin amplitudes ($S$) are realized under different external conditions. 
Cooperative changes in the magnetic states induced by the interacting spin-state degree of freedom are termed the ``spin-crossover'' or ``spin-state transition" phenomena, which are often seen in correlated electron materials,~\cite{imada, tokura, gretarsson} earth-inner mantels,~\cite{lin, antonangeli, hsu} biomolecules,~\cite{halder} and so on. 
These phenomena originate from competition between the crystalline electric-field (CEF) effect and the Hund coupling; the low (high)-spin state with small (large) $S$ is stabilized, when CEF is larger (smaller) than the Hund coupling.   

One of the well-known materials in which the spin-crossover phenomena are realized is the cobalt oxides with the perovskite structure, $R_{1-x}A_x$CoO$_3$ ($R$: trivalent ion, $A$: divalent ion).~\cite{mahendiran} 
The nominal valence of the Co ion at $x=0$ is $3+$ and the number of the $3d$ electrons is six. 
There are three possible spin states, the low-spin (LS) state with $S=0$ and the configuration of $(t_{2g})^6(e_g)^0$, the intermediate-spin state with $S=1$ and $(t_{2g})^5(e_g)^1$, and the high-spin (HS) state with $S=2$ and $(t_{2g})^4(e_g)^2$. 
When the LS (HS) states are realized in all cobalt sites in a crystal, the system is identified as a band (Mott) insulator. 
Thus, the spin-state transition phenomenon is considered as a phase transition between the band and Mott insulators. 
A number of the experimental observations for the spin crossover in the cobaltites have been reported over the past decades.~\cite{tokura, sato, vanko, a_ikeda} 
In recent years, studies of the cobaltites are revived from the viewpoint of the excitonic insulating (EI) state, which is defined as the quantum-mechanical hybridization of the different spin states.~\cite{kunes1, kunes2, kanamori, nasu, tatsuno} 

Recently, a multiferroic behavior was discovered in a distorted perovskite cobaltite BiCoO$_3$.~\cite{belik} 
Oka {\it et al.} found a ferroelectric (FE) transition and reduction of the FE transition temperature ($T$) by applying  pressure.~\cite{oka} 
In the FE phase, the CoO$_6$ octahedra are largely deformed; the coordination number of the cobalt ion is changed from six in the paraelectric (PE) phase to five in the FE phase. 
From the X-ray structural analyses and X-ray emission spectroscopy, it is suggested that the spin-state change from LS to HS occurs due to this large structural change. 
Thus, this material is expected to be a novel class of ME materials,~\cite{okimoto, oguchi, chen, chen2, solovyev, milosevic} in which the electric polarization couples with the spin-state degree of freedom, and is expected to open a new route to the multiferroics near the boundary between the band and Mott insulators. 

In this Letter, motivated by the experiments in BiCoO$_3$, we examine the magnetic and dielectric properties in the spin-state transition system. 
We introduce a strong coupling model, derived from the two-orbital Hubbard model, interacting with the FE-type lattice distortion, and investigate the ground-state and finite-$T$ properties using the mean-field approximation. 
We find that the LS-to-HS transition occurs with accompanying the FE transition at which the electronic entropy increases with decreasing temperature. 
The obtained phase diagram well reproduces the experimental pressure-temperature phase diagram. 
In the LS and PE phase close to the spin-state transition boundary, the electric polarization is induced by applying a magnetic field. 
A possible observation of this novel ME effect is discussed. 

We construct a model describing the correlated electrons coupled with the lattice distortion that breaks the inversion symmetry.
The model Hamiltonian is given by  
\begin{align}
{\cal H}={\cal H}_e+{\cal H}_l+{\cal H}_{el} ,
\label{eq:hamiltonian}
\end{align}
where the first and second terms represent the electronic and lattice parts, respectively, and the third term is for the electron lattice coupling. 
The electronic part ${\cal H}_e$ is derived from the extended two-orbital Hubbard model with CEF. 
This model is decomposed into the on-site interaction term and the inter-site electron hopping term as 
${\cal H}_{e0}={\cal H}_u+{\cal H}_t$ 
with 
\begin{align}
 {\cal H}_u&=\Delta\sum_i n_{i a} +U\sum_{i \eta}n_{i\eta\uparrow}n_{i\eta\downarrow}+U'\sum_{i} n_{ia}n_{ib}\nonumber\\
&+J\sum_{i\sigma\sigma'}c_{ia\sigma}^\dagger c_{ib\sigma'}^\dagger c_{ia\sigma'}c_{ib\sigma}
 +I\sum_{i\eta\neq \eta'}c_{i\eta\uparrow}^\dagger c_{i\eta \downarrow}^\dagger c_{i\eta'\downarrow}c_{i\eta'\uparrow} , 
\label{eq:Hu}
\end{align}
and 
\begin{align}
{\cal H}_t=-\sum_{\left < ij \right >} \sum_ {\eta \sigma}
t_\eta 
(c_{i\eta \sigma}^\dagger c_{j\eta \sigma}+{\rm H.c.}). 
\label{eq:Ht}
\end{align}
We define that $c^\dagger_{i \eta \sigma} \ (c_{i \eta \sigma}) $ is the creation (annihilation) operator for an electron with the orbital $\eta(=a, b)$, and the spin $\sigma(=\uparrow, \downarrow)$ at the site $i$. 
We introduce in ${\cal H}_u$ that $\Delta$, $U$, $U'$, $J$, and $I$ represent the energy difference between the orbitals, the intra-orbital Coulomb interaction, inter-orbital Coulomb interaction, the Hund coupling, and the pair hopping interaction, respectively.  
The electron hoppings in ${\cal H}_t$ are taken into account between the same kinds of orbitals in the nearest neighbor (NN) bonds. 
The average electron number per site is set to be two, and the two-dimensional square lattice is considered for simplicity. 

In the strong coupling limit when the electron-electron interactions and $\Delta$ are larger than $t_\eta$, the electron configurations at each site in the low energy section are restricted to the spin-singlet state with the $a^2b^0$ configuration mainly and the spin-triplet state with the $a^1b^1$ configuration. 
These are referred as the LS and HS states represented by $\ket{L}$ and $\ket{H_\Gamma}$ with $\Gamma \equiv S_z=(+1, 0, -1)$, respectively. 
In order to describe these local electronic states, we introduce the spin operator ${\bm S}_i$ with the amplitude $S=1$, the pseudo-spin operator for the spin-state degree of freedom as   
$\tau_\Gamma^{x}=\kets{L}\bra{H_\Gamma}+\kets{H_\Gamma}\bra{L}$, 
$\tau_\Gamma^{y}=i(\kets{L}\bra{H_\Gamma}-\kets{H_\Gamma}\bra{L})$, 
and 
$\tau_\Gamma^{z}=\kets{H_\Gamma}\bra{H_\Gamma}-\kets{L}\bra{L}$, 
and 
$\tau^\gamma=\sum_\Gamma \tau_{ \Gamma }^\gamma$ for $\gamma=(x, y, z)$. 
Details are presented in Ref.~[\citen{nasu}]. 
By using the second order perturbational calculations with respect to the hopping integrals, 
the electronic part ${\cal H}_{e0}$ is transformed to the effective model given by 
\begin{align}
{\cal H}_e&= -h_z\sum_i\tau_i^z+J_{z}\sum_{\means{ij}}\tau_i^z\tau_j^z
+J_s\sum_{\means{ij}}\bm{S}_i\cdot\bm{S}_j\nonumber\\
&-\sum_{\means{ij}} \sum_{\Gamma=(0, \pm 1)}
\left ( J_x \tau_{\Gamma i}^x\tau_{\Gamma j}^x+J_y\tau_{\Gamma i}^y\tau_{\Gamma j}^y \right )  , 
\label{eq:He}
\end{align}
where $h_z$, $J_s$, $J_\gamma$ are represented by the parameters introduced in Eqs.~(\ref{eq:Hu}) and (\ref{eq:Ht}). 
The first term originates from the energy difference between the two orbitals, and the second term brings about the attractive interaction between the LS and HS states. 
The last two terms induce the spontaneous hybridization of the LS and HS states, i.e. the EI state. 

As for the lattice part, we consider the local lattice displacement $q_i$ at site $i$, which breaks the space inversion symmetry, and its canonical momentum $p_i$.~\cite{blink} 
For simplicity, we assume that the displacements occur along a direction corresponding to the $c$ axis in the distorted perovskite crystal of BiCoO$_3$.  
The FE-type interaction is introduced between $q_i$s. 
The lattice part of the Hamiltonian is given as 
\begin{align}
{\cal H}_{l}=\sum_i \left [ \frac{\alpha}{2} p_i^2+V( q_i) \right ]
-\frac{1}{2}\sum_{i \ne j} v_{ij}q_i q_j , 
\label{eq:Hl}
\end{align}
where $p_i$ and $q_i$ are defined to be dimensionless. 
The anharmonic local potential at site $i$ is introduced as $V(q_i)=(\alpha/2)q_i^2+(\beta/4)q_i^4$ with $\alpha>0$ and $\beta>0$. 
The FE-type interaction between $q$'s is represented by a positive constant $v_0 = \sum_{j (\neq i)} v_{ij}$ in the mean-field approximation without setting of a detailed form of $v_{ij}$. 
Between the electron and lattice displacement, we introduce the local interaction which stabilizes the HS state associated with the FE order. 
Considering the space inversion symmetry of $\tau^z$, this interaction is represented as 
\begin{align}
{\cal H}_{el}=-g \sum_i \tau_i^z q_i^2 , 
\label{eq:Hel}
\end{align}
where $g$ is a positive couping constant. 

\begin{figure}[t]
\includegraphics[width=\columnwidth,clip]{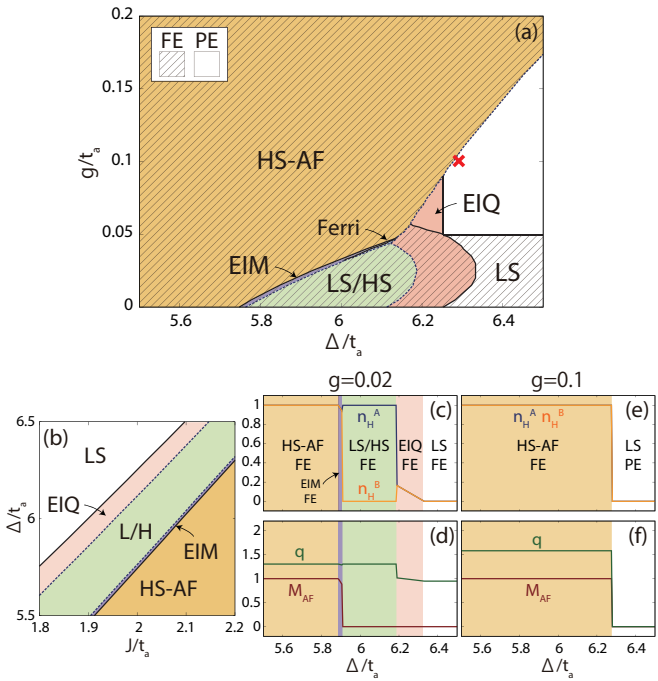}
\caption{(Color online) 
(a) 
Ground-state phase diagram in a $\Delta-g$ plane. 
Hatched areas represent the FE phases. 
(b) 
Ground-state phase diagram at $g=0$. 
Solid and broken lines represent the second- and first-order phase transitions, respectively. 
The $\Delta$ dependences of (c) $n_{\rm H}^A$ and $n_{\rm H}^B$, and (d) $q$ and $M_{\rm AF}$ at $g=0.02$.  
The results at $g=0.1$ are shown in (e) and (f). 
Other parameter values are chosen to be $t_a=1$, $t_b=-0.1$, $U=6J$, $U'=4J$, $J=I=2$, $\alpha=0.1$, $\beta=0.2$, and $v_0=0.4$. 
The red symbol at $(\Delta, g)=(6.28, 0.1)$ in (a) denotes the parameter value where the magnetic-field effect is examined.
}
\label{fig:fig1}
\end{figure}

The Hamiltonian in Eq.~(\ref{eq:hamiltonian}) is analyzed by using the mean-field approximation, where the interaction terms in ${\cal H}_e$ and ${\cal H}_l$ are decoupled as 
$\tau_i \tau_j \rightarrow \means{\tau_i} \tau_j+\tau_i \means{\tau_j}-\means{\tau_i}\means{\tau_j}$, 
${\bf S}_i {\bf S}_j \rightarrow \means{{\bf S}_i} {\bf S}_j+{\bf S}_i \means{{\bf S}_j}-\means{{\bf  S}_i}\means{{\bf S}_j}$,  and 
$q_i q_j \rightarrow \means{q_i} q_j+q_i \means{q_j}-\means{q_i}\means{q_j}$. 
We adopt a unit cell including the two sites, and calculate all the mean-field components selfconsistently. 
As for the lattice part, the FE-type order is assumed as $q = \langle q_i \rangle$. 
We introduce the AFM order parameter 
\begin{align}
M_{\rm AF}=\frac{1}{2} \left ( \means{S^z_A}-\means{S^z_B} \right ) , 
\end{align}
and the HS density
\begin{align}
n_{\rm H}= \sum_\Gamma \means { \ket{H_\Gamma}\bra{H_\Gamma} } , 
\end{align}
where subscripts $A$ and $B$ represent the two sublattices. 
In all of the numerical calculations except for Fig.~\ref{fig:fig1}(b) , the parameter values are fixed at $t_a=1$, $t_b=-0.1$, $U=6J$, $U'=4J$, $J=I=2$, $\alpha=0.1$, $\beta=0.2$, and $v_0=0.4$, and the orbital energy difference $\Delta$ and the electron-lattice coupling $g$ are changed. 

Before showing the numerical results of the the electron-lattice coupled system, the results at $g=0$ are briefly mentioned. Details are given in Ref.~[\citen{nasu}]. 
In Fig.~\ref{fig:fig1}(b), the ground-state phase diagram in the $J$-$\Delta$ plane at $g=0$ is presented. 
In the large limit of $\Delta$ ($J$), the LS (HS) phase appears as expected. 
An antiferromagnetic (AF) order is realized in the HS phase. 
Between the LS and HS phases, we find the two types of the EI phase, termed the EIM and EIQ phases, as well as the LS/HS ordered phase in which the LS and HS states are aligned alternately. 
The EI phases are identified by the order parameters $\means{\tau^x}$ and $\means{\tau^y}$, and 
the two phases  are distinguished by the magnetic structures: 
the AF order in EIM and the spin nematic order in EIQ. 
The LS/HS ordered phase is characterized by a staggered order of $\means{\tau^z}$. 

The ground-state phase diagram in the $\Delta$-$g$ plane is presented in Fig.~\ref{fig:fig1}(a). 
Hatched areas represent the FE phases. 
At $g=0$ where the electron and lattice degrees of freedom are independent with each other, the electronic states are changed as LS $\rightarrow$ EIQ $\rightarrow$ LS/HS $\rightarrow$ EIM $\rightarrow$ HS-AF with decreasing $\Delta$, and the FE-type lattice displacements occur. 
By introducing $g$ which promotes the HS state associated with the FE order, 
both the EIM and LS/HS phases are reduced and disappear around $g=0.04$. 
The phase is changed as $\rm LS \rightarrow EIQ \rightarrow HS$ with decreasing $\Delta$    
in the region of $0.04 < g<0.09$. 
With increasing $g$ furthermore, the FE HS phase and the PE LS phase only survive. 

Detailed $\Delta$ dependences of the mean fields, i.e. $q$, $M_{\rm AF}$, and $n_H$, are shown in Figs.~\ref{fig:fig1}(c)-\ref{fig:fig1}(f) at $g=0.02$ and $0.1$. 
The FE order parameter $q$ is finite except for the LS phase at $g=0.1$. 
The phase boundary between the HS-AF and EIM phases and that between the LS and EIQ phases are of the second order as shown in Ref.~[\citen{nasu}].  
Figures~\ref{fig:fig1}(e) and \ref{fig:fig1}(f) show that the transition between PE LS and FE HS-AF is of the first order. 

\begin{figure}[t]
\includegraphics[width=\columnwidth,clip]{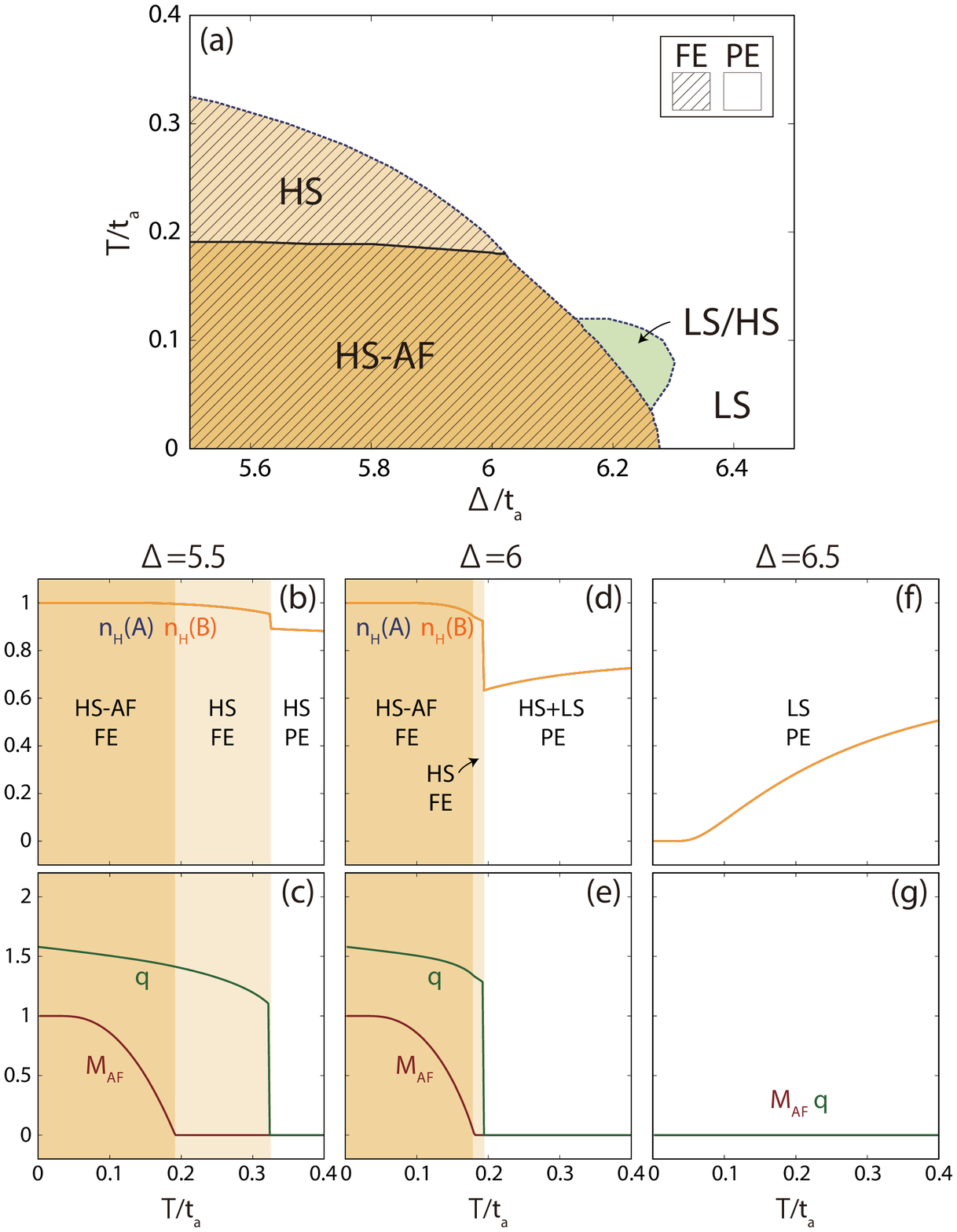}
\caption{(Color online) 
(a) Finite-$T$ phase diagram as a function of $\Delta$ at $g=0.1$. 
Hatched areas represent the FE states. 
Temperature dependences of $n_{\rm H}$, $q$, and $M_{\rm AF}$ at (b, c) $\Delta=5.5$, (d, e) $\Delta=6$, 
and (f, g) $\Delta=6.5$. 
Other parameter values are chosen to be $t_a=1$, $t_b=-0.1$, $U=6J$, $U'=4J$, $J=I=2$, $\alpha=0.1$, $\beta=0.2$, and $v_{0}=0.4$. 
}
\label{fig:fig2}
\end{figure}
%
The finite-$T$ phase diagram at $g=0.1$ is shown in Fig.~\ref{fig:fig2}(a). 
The temperature dependences of the mean fields at $\Delta=5.5$, $6.0$, $6.5$ are also shown in Figs.~\ref{fig:fig2}(b)-\ref{fig:fig2}(g). 
In high temperatures around $T=0.4$, with changing $\Delta$, the spin state is continuously changed, and there is no spin-state transition. 
In the large $\Delta$ region around $\Delta=6.5$,  $n_{\rm H}$ gradually decreases with decreasing $T$, and then arrives at $n_{\rm H}=0$, i.e. the LS state. 
On the other hand, in the small $\Delta$ region at $\Delta=5.5$, the PE-paramagnetic HS state changes into the FE-paramagnetic HS phase at $T \sim 0.33$, and the AF order appears around $T=0.19$. 
With increasing $\Delta$, the FE-transition temperature decreases monotonically, while the N\'{e}el temperature is almost unchanged. 
Around $\Delta=6.2$, the LS/HS ordered phase appears above the HS-AF phase. 
This is attributable to the spin entropy at the HS sites in the LS/HS phase, in which the exchange interactions between the HS sites are blocked by the surrounded LS sites. 
As shown in Figs.~\ref{fig:fig2}(d) and \ref{fig:fig2}(e) at $\Delta=6$, with decreasing $T$, $n_H$ decreases gradually, and increases discontinuously at the FE transition temperature. 
In the FE phase, $n_H$ increases monotonically with decreasing $T$. 
This finite-$T$ phase diagram as a function of $\Delta$ reproduces qualitatively the experimental phase diagram in BiCoO$_3$ under pressure, as discussed in more detail later.

\begin{figure}[t]
\includegraphics[width=\columnwidth,clip]{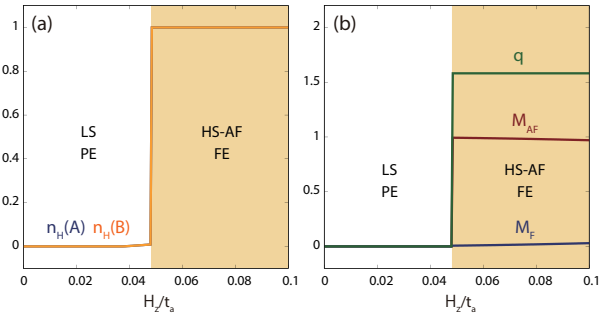}
\caption{(Color online) 
Magnetic-field dependences of (a) $n_{\rm H}$, (b) $M_{\rm AF}$, $M_{\rm F}$, and $q$  
at $T=0$, $g=0.1$, and $\Delta=6.28$, marked by the symbol in Fig.~\ref{fig:fig1}(a).  
Other parameter values are chosen to be $t_a=1$, $t_b=-0.1$, $U=6J$, $U'=4J$, $J=I=2$, $\alpha=0.1$, $\beta=0.2$, and $v_{0}=0.4$. 
}
\label{fig:fig3}
\end{figure}

Next, we show the ME effect in the present system. 
The magnetic field is introduced as the Zeeman energy into the Hamiltonian as ${\cal H}_{H}=- H \sum_i S_i^z$ with the magnetic field $H$. 
We examine the magnetic-field effect on a PE LS phase in the ground state marked by the symbol in Fig.~\ref{fig:fig1}(a).  
The magnetic field dependences of $n_{\rm H}$, $M_{\rm AF}$, $q$, and ferromagnetic order parameter $M_{\rm F}$ ($\equiv (\means{S_A^z}+\means{S_B^z})/2$) are presented in Fig.~\ref{fig:fig3}. 
With increasing $H$, the LS phase changes into the HS phase around $H=0.05$ accompanied by the FE distortion. 
That is, the magnetic-field-induced FE transition. 
Further increasing $H$, the spins are canted gradually and the angle between the sublattice magnetic moments deviates from 180$^\circ$. 
This nonlinear ME effect is owing to the coupling between the spin-state degree of freedom and the FE polarization.

\begin{figure}[t]
\includegraphics[width=\columnwidth,clip]{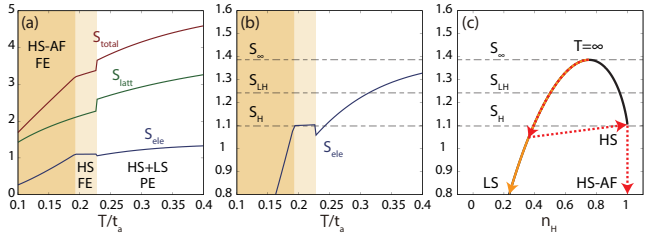}
\caption{(Color online) 
(a) Temperature dependences of the entropies. 
Red, blue, and green lines represent the total, electronic, and lattice parts of the entropies, respectively. 
(b) An enlarged figure of the temperature dependence of $S_{\rm ele}$.
Parameter values are chosen to be $t_a=1$, $t_b=-0.1$, $U=6J$, $U'=4J$, $J=I=2$, $\alpha=0.1$, $\beta=0.2$, $v_{0}=0.4$, $\Delta=6.5$, and $g=0.19$. 
(c) Schematic entropy change and a phase transition pass in the spin-state transition accompanied by the FE transition.  
Broken and solid arrows represent the entropy changes in the present system and in the conventional spin-crossover systems, respectively. 
}
\label{fig:fig4}
\end{figure}
%
We mention the entropy change across the FE transition. 
As shown schematically in Fig.~\ref{fig:fig4}(c), the present system is in stark contrast to the conventional spin-crossover system, in which a large spin entropy in the high temperature HS phase is quenched in the low temperature LS phase. 
The total entropy ($S_{\rm total}$) as well as the electronic ($S_{\rm ele}$) and lattice ($S_{\rm latt}$) entropies calculated by the mean-field approximation are shown in Fig.~\ref{fig:fig4}(a). 
Here, we chose $(\Delta, g)=(6.5, 0.19)$ in which the first-order FE transition accompanied by the spin-state change occurs around $T=0.23$, and the AF transition follows up around $T=0.2$. 
The detailed $T$ dependence of $S_{\rm ele}$ is presented in Fig.~\ref{fig:fig4}(b). 
The notable feature is that $S_{\rm ele}$ increases with decreasing $T$ at the FE transition temperature. 
The electronic part of the entropy is evaluated by the simple analytical formula  
$S=-n_{\rm L} \ln (n_{L}) - n_{\rm H} \ln (n_{\rm H}/3)$ with $n_{\rm L} = 1-n_{\rm H}$ where the LS and paramagnetic HS states are assumed to be mixed in the thermodynamic sense. 
This equation gives
$S=\ln 4 \sim 1.39 (\equiv S_\infty )$ in the high $T$ limit with $(n_{\rm H}, n_{\rm L})=(0.75, 0.25)$, 
 $S=0.5 \ln 2+0.5\ln 6 \sim 1.24 (\equiv S_{\rm LH})$ in the case with the equal weight of the LS and HS states of $(n_{\rm H}, n_{\rm L})=(0.5, 0.5)$, 
and $S=\ln 3 \sim 1.09(\equiv S_{\rm H})$ in the pure HS state of $(n_{\rm H}, n_{\rm L})=(1, 0)$.  
As shown in Fig.~\ref{fig:fig4}(b), the calculated $S_{\rm ele}$ approaches $S_{\infty}$ in the high $T$ limit, and $S_{\rm H}$ in the FE HS phase. 
Between $T = 0.23$ and $0.3$, $S_{\rm ele}$ is smaller than $S_{\rm LH}$.
These results imply that with decreasing $T$ the electronic system once approaches the LS state and then changes into the HS state owing to the FE transition. 
This is schematically shown in Fig.~\ref{fig:fig4}(c), and  is an opposite behavior of a conventional spin-crossover system such as LaCoO$_3$.

Finally, relations of the present calculations to the experimental observations are discussed. 
The $\Delta$-$T$ phase diagram shown in Fig.~\ref{fig:fig2}(a) is compared to the pressure-$T$ phase diagram in BiCoO$_3$~\cite{oka}. 
The hydrostatic pressure reduces the lattice constants and increases CEF. 
Therefore, the calculated reduction of the FE transition temperature is consistent qualitatively with the experimental results. 
We predict that, in contrast to the FE transition temperature, the AF N\'{e}el temperature is almost unchanged under the pressure, because the HS density $n_{\rm H}$ is almost independent of $\Delta$ in the FE phase. 
One of the present main message is that the nonlinear ME effect, i.e. the magnetic-field-induced electric polarization, is expected in the PE LS phase close to the FE HS phase associated with the AF order. 
This might be confirmed experimentally in BiCoO$_3$ under the high pressure around $3$ GPa below $420$ K (the N\'{e}el temperature at ambient pressure).~\cite{belik, oka} 
The present theoretical prediction of the ME effect is also checked in the mixed crystal of BiCo$_{1-x}$Fe$_x$O$_3$ where the fine tuning of the FE transition is possible.~\cite{hojo} 

In conclusion, we study the mangetoelectric effects in the spin-state transition system coupled with the ferroelectric lattice distortion. The ferroelectric transition occurs accompanied by the LS-to-HS transition. 
The calculated $\Delta$-$T$ phase diagram is consistent with the experimental pressure-$T$ phase diagram observed  in BiCoO$_3$. 
It is predicted that the electric polarization is induced by applying the magnetic field in the LS paraelectric phase. 
The present study proposes a concept of the new-class multiferroics in the spin-state transition system with electron correlations. 

We thank M. Azuma for their helpful discussions.
This work was supported by MEXT KAKENHI Grant No. JP26287070, JP15H02100, JP16H00987, JP16K17747, and JP16K17731. 
Some of the numerical calculations were performed using the facilities of the Supercomputer Center, the Institute for Solid State Physics, the University of Tokyo.

\end{document}